\documentclass[11pt]{article}

\usepackage[margin=1in]{geometry}
\usepackage{amsmath,amssymb,amsfonts}
\usepackage{setspace}
\usepackage{authblk}
\usepackage{graphicx}
\usepackage{subcaption}
\usepackage{xcolor}
\usepackage{caption}
\usepackage{booktabs}
\usepackage{array}
\usepackage{parskip}

\title{Semiparametric Models for Practice Effects in Longitudinal Cognitive Trajectories:\\
Application to an Aging Cohort Study}

\author[1]{Yunshan Xu\thanks{Corresponding author: yunshanxu@ucsd.edu}}
\author[1]{Tsungchin Wu}
\author[2]{Angelina Van Dyne}
\author[2]{Ellen Lee}
\author[2]{Lisa Eyler}
\author[1]{Xin M.\ Tu}

\affil[1]{Herbert Wertheim School of Public Health and Human Longevity Science,\\
University of California San Diego, La Jolla, CA, USA}
\affil[2]{Department of Psychiatry, University of California San Diego,\\
La Jolla, CA, USA}

\date{} 

\begin{document}

\maketitle
\begin{abstract}
\noindent
\textbf{Background:} True cognitive longitudinal decline can be obscured by repeated testing, which is called practice effects (PEs). We developed a modeling framework that aligns participants by baseline and estimates visit-specific PEs independently of age-related change.

\noindent
\textbf{Method:} Using real data ($N=175$), we estimated within-subject correlations via linear mixed-effects modeling and applied these parameters to simulate longitudinal trajectories for healthy controls (HC) and individuals with schizophrenia (SZ). Simulations incorporated aging, diagnostic differences, and cumulative PE indicators. Generalized estimating equations (GEEs) were fit with and without PEs to compare model performance.

\noindent
\textbf{Results:} Models that ignored PEs inflated estimates of cognitive stability and attenuated HC--SZ group differences. Including visit-specific PEs improved recovery of true trajectories and more accurately distinguished aging effects from learning-related gains. Interaction models further identified that PEs may differ by diagnosis or by age at baseline.

\noindent
\textbf{Conclusion:} Practice effects meaningfully bias longitudinal estimates if left unmodeled. The proposed alignment-based GEE framework provides a principled method to estimate PEs and improves accuracy in both simulated and real-world settings.

\noindent
\textbf{Keywords:} practice effects; repeat testing; serial testing; longitudinal testing; mild cognitive impairment; cognitive change.
\end{abstract}
\newpage
\section{Introduction}
Longitudinal studies are crucial for understanding the course of cognitive impairment in schizophrenia, including whether deficits are static, progressive, or partially reversible with treatment~\cite{Szoeke2008,Fioravanti2012,Rund1998}. In clinical studies of cognitive impairments associated with schizophrenia, repeated neuropsychological testing is a key outcome measure used to evaluate both pharmacological and psychosocial interventions. However, it has been shown that improvements in test performance over time are more likely to be due to practice effects (PEs) than to genuine changes in underlying abilities~\cite{Goldberg2010,Keefe2017,Hill2010}. PEs can therefore distort conclusions about the natural course of schizophrenia and falsify the assessment of a treatment's effectiveness, especially in studies that use changes in cognitive scores as the primary endpoint.
\\

A substantial body of literature has now confirmed the existence of a practice effect in schizophrenia patients undergoing various cognitive tests. Goldberg's review indicates that cognitive improvements from targeted training are often underestimated due to confounding factors, and many findings regarding “cognitive enhancement” from antipsychotic drugs can be reinterpreted as training effects or placebo effects~\cite{Goldberg2010,Goldberg2007}. In a study of second-generation antipsychotics in first-episode schizophrenia patients, he noted that the observed cognitive improvement was comparable to the practice effect seen in healthy controls, indicating minimal drug-specific cognitive effects~\cite{Goldberg2007}. Meanwhile, Keefe et al. analyzed data from 12 placebo-controlled cognitive studies and found that training effects significantly influenced MATRICS Consensus Cognitive Scale scores, with improvements comparable to those in the placebo group~\cite{Keefe2017}. Collectively, these studies indicate that failing to account for training effects may lead to overestimation of cognitive improvement and misjudgment of the longitudinal stability or improvement of cognitive function in schizophrenia patients.
\\

Practice effects (PEs) are sometimes operationalized as the degree of improvement from baseline to follow-up, but may also manifest as a reduction in the rate of decline. Longitudinal meta-analyses indicate that neuropsychological performance in individuals with schizophrenia often shows only modest improvement or relative stability, even in the presence of expected disease-related and age-related decline~\cite{Szoeke2008,Fioravanti2012,Rund1998}. This pattern suggests that practice and related nonspecific factors may mask underlying deterioration, particularly when follow-up intervals are short or subjects undergo multiple testing. For example, Granholm et al. found no neurodegenerative decline in neuropsychological tests administered over multiple years to middle-aged and elderly schizophrenia patients; conversely, aging positively correlated with diminishing gains on repeated tasks, suggesting an interaction between aging, baseline impairment, and practice effects~\cite{Granholm2010}. These findings underscore that practice effects may persist even when observed scores plateau or decline, with their intensity varying across individuals.
\\

Despite this evidence, explicit adjustment for PEs is not a standard approach in schizophrenia research. Many longitudinal and trial analyses implicitly assume that changes of scores reflect the true cognitive change, with practice treated as noise or ignored. There has been relatively limited work modeling visit-specific PEs, examining how PEs evolve across multiple assessments, or testing whether PEs differ by diagnosis or age. Most existing studies focus on short test-retest intervals (weeks to months) rather than multi-year follow-up, and few integrate practice-effect modeling into marginal models of cognitive trajectories in schizophrenia.
\\

In the present work, we address these gaps by developing a generalized estimating equation (GEE) framework that aligns participants by baseline assessment, treats age as a continuous time scale, and incorporates visit-specific indicators to estimate practice effects separately from age-related change. Using both real and simulated data on individuals with schizophrenia and healthy controls, we evaluate how PEs accumulate over repeated assessments, whether their magnitude varies by age or diagnosis, and how failure to model PEs biases estimates of cognitive trajectories. Our overarching goal is to provide a practical approach for incorporating practice effects into longitudinal cognitive analyses in schizophrenia, thereby improving the validity of inferences about disease course and treatment response.

\section{Methods}

\subsection{Study Sample}
Participants were drawn from an ongoing longitudinal study of psychosis and aging conducted at the University of California San Diego. The study follows adults with schizophrenia (SZ) and demographically similar healthy controls (HC) to characterize cognitive aging across midlife and later adulthood. Individuals with SZ were recruited through outpatient psychiatric clinics and community mental health programs, whereas HC participants were recruited through community advertisements and referral networks. All participants were fluent in English and had sufficient sensory and motor ability to complete neuropsychological testing.
\\

At baseline, 175 participants completed cognitive assessments, including  90 SZ and 85 HC, and were between 26.2–49.8 years of age (mean = 39.2, SD = 7.0). Participants returned for follow-up evaluations approximately every 12 months, with up to six repeated assessments per individual. Demographic characteristics, psychiatric diagnoses, medication history, and medical comorbidities were collected at each visit. Cognitive performance was assessed using a standardized battery, and a composite score was derived to summarize global cognitive functioning.
Participants were excluded if they had a neurological disorder, intellectual disability, uncontrolled medical illness affecting cognition, or recent substance dependence. The analytic sample included individuals with at least one post-baseline cognitive assessment. All procedures were approved by the UC San Diego Human Research Protections Program (IRB Protocol $\#$101631), and written informed consent was obtained from all participants.

\subsection{Statistical Analysis}

\subsubsection*{Alignment of Assessment Time to Estimate Practice Effects}
In this study, subjects are assessed approximately every 12 months (there is some
deviation for some subjects, but we treat assessments as 12 months apart for
analytic purposes). Thus, we can estimate practice effects (PEs) for the 1st,
2nd, 3rd, etc.\ reassessments. For simplicity, we focus on the practice effect
for the 1st reassessment.

Because there are 12 months between consecutive assessments, we can estimate the
practice effect for the 1st reassessment at 12 months. We partition the sample
into the following subsamples.

\begin{figure}[ht]
\centering

\begin{minipage}{0.45\textwidth}
\centering
\caption*{Age at visit for each observed subject}
\begin{tabular}{ccc}
\toprule
Age at enrollment & 1st visit & 2nd visit \\
\midrule
26 & 26 & 27 \\
27 & 27 & 28 \\
28 & 28 & 29 \\
29 & 29 & 30 \\
\vdots & \vdots & \vdots \\
\bottomrule
\end{tabular}
\end{minipage}
\hfill
\begin{minipage}{0.45\textwidth}
\centering
\caption*{PE for each age aligned on 1\textsuperscript{st} reassessment}
\begin{tabular}{cccccc}
\toprule
 & & & & & \\
26 & \textcolor{red}{27} & & & & \\
 & 27 & \textcolor{red}{28} & & & \\
 & & 28 & \textcolor{red}{29} & & \\
 & & & 29 & \textcolor{red}{30} & \\
\vdots & \vdots & \vdots & \vdots & \vdots & \\
\bottomrule
\end{tabular}
\end{minipage}

\caption{Left: observed ages at enrollment and first two study visits. 
Right: practice effect illustration obtained by aligning assessments on the first reassessment; red values represent performance at the reassessment.}
\end{figure}

We model practice effects by examining the difference in mean between the red
values (2nd assessment, or 1st reassessment) and the black values (1st
assessment). By including an interaction between practice effect and age at
visit, we can also test whether practice effects change as age increases.

We dummy-code age at visit and practice-effect indicators so that we do not
impose a specific functional form on the relationship between the cognitive
outcome, age at visit, and practice effect. We first included age at visit as
28, 29, 30, etc.\ as single-year indicators; because results were unstable, we
then grouped age at visit into 5-year bins so that the model provides the mean
cognitive outcome for each age bin and a practice effect for each age bin. The
results for \texttt{EXCOMP2} showed that, without controlling for practice
effects, \texttt{EXCOMP2} appeared to increase over age, but after adjusting for
practice effects, the mean of \texttt{EXCOMP2} decreased with age.

\subsubsection*{Generalized Estimating Equation Models}

\textbf{Models Without Practice Effects}

We first estimated a GEE model that omitted explicit practice effects. Let $i$ index subjects and $j$ index assessment times. Let $Y_{ij}$ denote the observed response (dependent variable) and $W_{ij}$ the vector of all explanatory variables, including age at visit, diagnostic group $dx_i$, and other baseline covariates (e.g., gender, education, race). Time since baseline is represented by $t_{ij}$ in years, where $j$ indexes the $j$th visit and $t_{i1}=0$ at baseline. Age at visit, $\text{agevisit}_{ij}$, is defined as the subject $i$'s age at visit $j$, and $\text{age}_{i1}$ is the age of subject $i$ at the baseline visit.

The model without PEs can be written as
\begin{equation}
\mathbb{E}[Y_{ij}\mid W_{ij}]=
\beta_0 + \beta_1 \text{agevisit}_{ij}
+ \beta_2 dx_i
+ \beta_3 t_{ij} dx_i
+ X_i^\top \beta_4,
\end{equation}
which is equivalent to
\begin{equation}
\mathbb{E}[Y_{ij}\mid W_{ij}] =
\beta_0 + \beta_1 (\text{age}_{i1}+t_{ij})
+ \beta_2 dx_i
+ \beta_3 t_{ij} dx_i
+ X_i^\top \beta_4.
\end{equation}

This model serves as the baseline against which PE-adjusted models are compared, allowing us to quantify the extent to which ignoring repeated-test gains may bias longitudinal estimates.

\textbf{Models With Practice Effects}

To estimate PEs explicitly, we extended the GEE model by introducing visit-specific indicator variables. Let $I(j=k)$ be an indicator function that equals 1 if visit $j$ corresponds to the $k$th assessment and 0 otherwise. Incorporating these indicators, the model becomes
\begin{equation}
\begin{split}
\mathbb{E}[Y_{ij}\mid W_{ij}] &=
\beta_0 + \beta_1 \text{agevisit}_{ij}
+ \beta_2 dx_i
+ \beta_3 t_{ij} dx_i
+ X_i^\top \beta_4
+ \sum_{k=2}^{5} \beta_{5k} I(j=k) \\
&\quad + \sum_{k=2}^{5} \beta_{6k} I(j=k) dx_i,
\end{split}
\end{equation}
or equivalently,
\begin{equation}
\begin{split}
\mathbb{E}[Y_{ij}\mid W_{ij}] &=
\beta_0 + \beta_1 (\text{age}_{i1}+t_{ij})
+ \beta_2 dx_i
+ \beta_3 t_{ij} dx_i
+ X_i^\top \beta_4
+ \sum_{k=2}^{5} \beta_{5k} I(j=k) \\
&\quad + \sum_{k=2}^{5} \beta_{6k} I(j=k) dx_i.
\end{split}
\end{equation}
Here, $\beta_{5k}$ captures the PE at visit $k$ ($2\le k\le 5$), and the interaction coefficients $\beta_{6k}$ capture group differences in PEs between healthy controls and individuals with schizophrenia.

\subsection{Simulation Study}

A simulation study was conducted to evaluate the performance of the aligned GEE framework under controlled conditions and to examine how failing to model PEs can bias estimates of longitudinal cognitive trajectories.

\subsubsection{Estimating the Correlation Structure}

We first estimated the within-subject correlation structure from real longitudinal cognitive data. Participants with at least one post-baseline visit were retained, and up to six visits per subject were used. Time since baseline was derived from days-since-baseline and converted to years. Subjects without a baseline assessment were excluded to maintain proper alignment. A linear mixed-effects model was fit to these data to estimate the intra-class correlation (ICC):
\begin{equation}
Y_{ij} = \beta_0 + \beta_1 \text{agevisit}_{ij} + \beta_2 dx_i + \beta_3 t_{ij} dx_i + X_i^\top \beta_4 + b_i + \epsilon_{ij},
\end{equation}
\begin{equation}
b_i \sim \mathcal{N}(0,\sigma_b^2),\quad \epsilon_{ij} \sim \mathcal{N}(0,\sigma_e^2),
\end{equation}
or equivalently,
\begin{equation}
Y_{ij} = \beta_0 + \beta_1 (\text{age}_{i1}+t_{ij}) + \beta_2 dx_i + \beta_3 t_{ij} dx_i + X_i^\top \beta_4 + b_i + \epsilon_{ij},
\end{equation}
\begin{equation}
b_i \sim \mathcal{N}(0,\sigma_b^2),\quad \epsilon_{ij} \sim \mathcal{N}(0,\sigma_e^2).
\end{equation}
Here $b_i$ and $\epsilon_{ij}$ denote the subject-specific random intercept and residual error, respectively. The fitted model produced the following parameter estimates:
\[
\beta_0 = 0.333,\quad \beta_1 = -0.006,\quad \beta_2 = -0.803,\quad \beta_3 = 0.018,
\]
\[
\beta_4 = [0.093,\,0.091,\,-0.078]^\top,\quad \sigma^2 = (0.312)^2,\quad \rho = 0.753.
\]
The ICC was calculated as
\[
\text{ICC} = \frac{\sigma_b^2}{\sigma_b^2 + \sigma_e^2},
\]
and this value was subsequently used to calibrate the copula-based dependence structure in the simulated data.

Using both real and simulated datasets, we examined longitudinal outcome trajectories for individuals in the Healthy Control (HC) and Schizophrenia (SZ) groups, comparing scenarios that included practice effects with those that did not. To generate within-subject dependence that better resembled real cognitive data, we employed a copula-based approach to model the correlation structure. Marginal mean trajectories were then estimated with generalized estimating equations (GEE), which provided a population-level view of how outcomes evolved over time under each modeling condition.

\subsubsection{Simulation With No Practice Effects (No PE)}

To create a more realistic scenario and decouple practice effects from time, we simulated data for 
$n = 500$ subjects in five waves of enrollment. Let $t_{ij}$ denote the time in years at the $j$th visit 
with $t_{i1} = 0$ $(1 \le i \le n,\; 1 \le j \le J)$. For our simulation study, we set
\[
\beta_{0} = 0.326,\quad 
\beta_{1} = -0.007,\quad 
\beta_{2} = -0.782,\quad 
\beta_{3} = 0.013,\quad 
\boldsymbol{\beta}_{4} = [0.098,\; 0.034,\; -0.077],
\]
\[
n = 500,\quad J = 5,\quad t_{ij} = 1 \times (j-1),\quad \sigma^{2} = (0.155)^{2}.
\]

Thus the five assessment times occurred at 
\[
t_{i1}=0,\quad 
t_{i2}=1,\quad 
t_{i3}=2,\quad 
t_{i4}=3,\quad 
t_{i5}=4,\quad 
t_{i6}=5.
\]

A new cohort of 100 individuals of the same age in months was then randomly assigned to either the 
schizophrenia (SZ) or healthy control (HC) group, with their first assessment at $j=1$ and $t_{i1}=0$. 
The outcome $Y_{ij}$ for subject $i$ at visit $j$ was generated as:
\begin{equation}
Y_{ij} = \mu_{ij} + \epsilon_{ij}, \qquad 
\mu_{ij} = \beta_{0} + \beta_{1}\,\text{agevisit}_{ij} + \beta_{2} d_{x_i} 
+ \beta_{3} t_{ij} \, d_{x_i} + \mathbf{X}_i^{\top} \boldsymbol{\beta}_{4},
\end{equation}
or equivalently,
\[
\mu_{ij} = \beta_{0} + \beta_{1}\left(\text{age}_{i1} + t_{ij}\right)
+ \beta_{2} d_{x_i} + \beta_{3} t_{ij} \, d_{x_i} 
+ \mathbf{X}_i^{\top} \boldsymbol{\beta}_{4},
\]
where $d_{x_i} = 1$ for SZ and $0$ for HC, and $\text{age}_{i1}$ denotes the age (in years) of the $i$th 
subject at baseline ($j = 1$, $t_{i1}=0$). We set $\text{age}_{i1}$ for each cohort as:
\[
\text{age}_{i1} =
\begin{cases}
25 & \text{Cohort 1},\\
30 & \text{Cohort 2},\\
35 & \text{Cohort 3},\\
40 & \text{Cohort 4},\\
45 & \text{Cohort 5}.
\end{cases}
\]

In model (1), 
\[
\text{agevisit}_{ij} = \text{age}_{i1} + t_{ij}
\]
is the age of subject $i$ at visit $j$, and $\beta_{1}$ is the aging effect. 
We can also model aging effects as a categorical variable using:
\begin{equation}
Y_{ij} = \mu_{ij} + \epsilon_{ij}, \qquad 
\mu_{ij} = \beta_{0} + \sum_{k=1}^{6} 
\mathbf{1}\!\left(5(k-1) + 25 \le \text{agevisit}_{ij} \le 5k + 25\right)\beta_{1} 
+ \beta_{2} d_{x_i} + \beta_{3} t_{ij} d_{x_i} + \mathbf{X}_i^{\top}\boldsymbol{\beta}_{4},
\end{equation}

\subsubsection{Simulation With Practice Effects (PE)}

To incorporate cumulative practice effects, we extended the simulation framework by adding 
five indicator variables, \texttt{prac1plus} through \texttt{prac5plus}, denoting whether a 
subject had completed at least one through five previous assessments. These indicators were 
included as main effects and in interaction with diagnosis.

The linear predictor for subject $i$ at visit $j$ was specified as:
\[
Y_{ij} = \mu_{ij} + \epsilon_{ij},
\]
with
\[
\mu_{ij} = 
\beta_0 
+ \beta_1 \text{agevisit}_{ij}
+ \beta_2 d_i
+ \beta_3 t_{ij} d_i
+ \mathbf{X}_i^{\top}\beta_4
+ \sum_{k=2}^{6} \beta_{5k} I(j = k),
\]
or equivalently,
\[
\mu_{ij} = 
\beta_0 
+ \beta_1(a\!ge_{i1} + t_j)
+ \beta_2 d_i
+ \beta_3 t_{ij} d_i
+ \mathbf{X}_i^{\top}\beta_4
+ \sum_{k=2}^{6} \beta_{5k} I(j=k).
\]

Here, $\beta_{5k}$ represents the increase in performance due to practice at visit $k$, with
\[
\beta_{52} \le \beta_{53} \le \beta_{54} \le \beta_{55} = \beta_{56},
\]
and we set
\[
\beta_{52} = 0.2,\quad 
\beta_{53} = 0.3,\quad 
\beta_{54} = 0.4,\quad 
\beta_{55} = \beta_{56} = 0.5.
\]

\textbf*{Differential Practice Effects by Diagnosis}

To evaluate group differences in practice effects between schizophrenia (SZ) and healthy 
control (HC) participants, we used:
\[
Y_{ij} = \mu_{ij} + \epsilon_{ij},
\]
\[
\mu_{ij} =
\beta_0
+ \beta_1 \text{agevisit}_{ij}
+ \beta_2 d_i
+ \beta_3 t_{ij} d_i
+ \mathbf{X}_i^{\top}\beta_4
+ \sum_{k=2}^{5} \beta_{5k} I(j=k)
+ \sum_{k=2}^{5} \beta_{6k} I(j=k)d_i,
\]
or
\[
\mu_{ij} =
\beta_0
+ \beta_1(a\!ge_{i1} + t_j)
+ \beta_2 d_i
+ \beta_3 t_{ij} d_i
+ \mathbf{X}_i^{\top}\beta_4
+ \sum_{k=2}^{5} \beta_{5k} I(j=k)
+ \sum_{k=2}^{5} \beta_{6k} I(j=k)d_i.
\]

Here, $\beta_{6k}$ represents the differential practice effect for SZ relative to HC at visit $k$.

\textbf*{Differential Practice Effects by Age}

To evaluate whether practice effects vary by age, we used:
\[
Y_{ij} = \mu_{ij} + \epsilon_{ij},
\]
\[
\mu_{ij} =
\beta_0
+ \beta_1 \text{agevisit}_{ij}
+ \beta_2 d_i
+ \beta_3 t_{ij} d_i
+ \mathbf{X}_i^{\top}\beta_4
+ \sum_{k=2}^{5} \beta_{5k} I(j=k)
+ \sum_{k=2}^{5} \beta_{7k} I(j=k)\,\text{agevisit}_{ij},
\]
or
\[
\mu_{ij} =
\beta_0
+ \beta_1(a\!ge_{i1} + t_j)
+ \beta_2 d_i
+ \beta_3 t_{ij} d_i
+ \mathbf{X}_i^{\top}\beta_4
+ \sum_{k=2}^{5} \beta_{5k} I(j=k)
+ \sum_{k=2}^{5} \beta_{7k} I(j=k)(a\!ge_{i1} + t_j).
\]

Here, $\beta_{7k}$ quantifies the change in practice-related improvement per additional year 
of age at visit $k$. As in earlier sections, aging effects can alternatively be modeled using 
categorical age groups.

\section{Results}

Baseline demographic characteristics were comparable between groups (Table 1). The schizophrenia group included a slightly higher proportion of males (53\% vs. 45\% in controls) and had fewer years of education (median 12 vs. 15). Median baseline age was similar across groups—40 years for schizophrenia and 38 years for healthy controls. Racial/ethnic distributions did not differ substantially, with most participants identifying as Caucasian, followed by Hispanic and African American.
\begin{table}[ht]
\centering
\caption{Baseline characteristics of study participants}
\label{tab:baseline}
\begin{tabular}{lcc}
\toprule
\textbf{Characteristic} & \textbf{Healthy controls (n = 85)} & \textbf{Schizophrenia (n = 90)} \\
\midrule
\textbf{Sex}, male, n (\%) & 38 (45\%) & 48 (53\%) \\
\textbf{Age at baseline}, median (Q1, Q3), years & 38 (32, 46) & 40 (34, 46) \\
\textbf{Years of education}, median (Q1, Q3) & 15 (13, 16) & 12 (11, 13) \\
\midrule
\multicolumn{3}{l}{\textbf{Race/ethnicity}, n (\%)} \\
\quad Caucasian & 49 (58\%) & 38 (42\%) \\
\quad African American & 11 (13\%) & 12 (13\%) \\
\quad Hispanic & 21 (25\%) & 31 (34\%) \\
\quad Asian & 3 (3.5\%) & 4 (4.4\%) \\
\quad Native Hawaiian / Pacific Islander & 0 (0\%) & 1 (1.1\%) \\
\quad Multiracial & 1 (1.2\%) & 4 (4.4\%) \\
\bottomrule
\end{tabular}
\end{table}

Table 2 summarizes the linear mixed-effects model fit to the
observed data. Cognitive performance was substantially lower in SZ compared with
HC ($\beta = -0.803$, $p < 0.0001$). Higher educational attainment was associated
with better performance ($\beta = 0.093$, $p = 0.0001$), whereas age at visit was
not statistically significant ($p = 0.34$). A significant diagnosis-by-time
interaction ($\beta = 0.018$, $p = 0.0037$) indicated differing longitudinal
trajectories between groups.
\\

To evaluate whether these estimates reflected practice effects, simulated
datasets were generated using the real-data parameter values but without
practice effects. As shown in Tables 3 and 4, the schizophrenia effect remained similar
($\beta \approx -0.78$, $p < 2\times10^{-16}$), and age demonstrated a small,
negative slope ($\beta \approx -0.007$, $p < 0.001$), confirming that the
simulation accurately reproduced the empirical effect structure.
\\

Table 6 presents GEE estimates from simulations incorporating
practice effects. Large gains were observed at early reassessments (e.g.,
prac1: $\beta = 0.305$, $p < 2\times10^{-16}$), with diminishing increases
thereafter. When practice effects were modeled, the expected age-related decline
re-emerged ($\beta = -0.006$, $p = 0.0006$), demonstrating that unmodeled
practice effects can mask true cognitive deterioration.

Figures 2 and 3 illustrate these patterns: parameter estimates were consistent across analytic frameworks, and trajectories that appeared stable without adjustment showed clear decline once practice effects were accounted for.

\begin{table}[hbtp]
\caption{Linear Mixed Effects Model for Correlation (Real Data)}%
\centering
\begin{tabular}{lrrrrr}
\hline
\textbf{Fixed Effect} & \textbf{Estimate} & \textbf{Std. Error} & \textbf{DF} & \textbf{t-value} & \textbf{p-value} \\
\hline
(Intercept)         & 0.333  & 0.515 & 397 & 0.65  & 0.5177 \\
agevisit            & -0.006 & 0.006 & 397 & -0.95 & 0.3424 \\
dxgroup             & -0.803 & 0.109 & 170 & -7.39 & $< 0.0001$ \\
educ                & 0.093  & 0.022 & 170 & 4.16  & 0.0001 \\
gender              & 0.091  & 0.089 & 170 & 1.02  & 0.3071 \\
race\_lat           & -0.078 & 0.034 & 170 & -2.30 & 0.0228 \\
dxgroup:yearsbl     & 0.018  & 0.006 & 397 & 2.92  & 0.0037 \\
\hline
\end{tabular}
\end{table}

\begin{table}[hbtp]
\caption{Linear Mixed Effects Model (Simulated Data, Without Practice Effect)}%
\centering
\begin{tabular}{lrrrrr}
\hline
\textbf{Fixed Effect} & \textbf{Estimate} & \textbf{Std. Error} & \textbf{DF} & \textbf{t-value} & \textbf{p-value} \\
\hline
(Intercept)      & 0.319 & 0.0932 & 1519 & 3.42  & 0.0006 \\
age\_visit       & -0.007 & 0.0015 & 1519 & -4.79 & 0.0000 \\
dx\_bin          & -0.782 & 0.0252 & 495  & -30.99 & 0.0000 \\
educ             & 0.099  & 0.0049 & 495  & 20.11 & 0.0000 \\
gen              & 0.035  & 0.0247 & 495  & 1.40  & 0.1632 \\
race\_lat        & -0.077 & 0.0092 & 495  & -8.29 & 0.0000 \\
dx\_bin:t        & 0.013  & 0.0035 & 1519 & 3.71  & 0.0002 \\
\hline
\end{tabular}
\end{table}

\begin{table}[hbtp]
\caption{GEE Results (Simulated Data, Without Practice Effect)}%
\centering
\begin{tabular}{lrrrr}
\hline
\textbf{Term} & \textbf{Estimate} & \textbf{Std. Err} & \textbf{Wald} & \textbf{p-value} \\
\hline
(Intercept)     & 0.32629 & 0.09170 & 12.66  & 0.00037 \\
age\_visit      & -0.00754 & 0.00150 & 25.19  & $5.2 \times 10^{-7}$ \\
dx\_bin         & -0.78211 & 0.02510 & 970.66 & $< 2 \times 10^{-16}$ \\
educ            & 0.09886  & 0.00499 & 390.98 & $< 2 \times 10^{-16}$ \\
gen             & 0.03441  & 0.02450 & 1.97   & 0.16014 \\
race\_lat       & -0.07660 & 0.00790 & 94.10  & $< 2 \times 10^{-16}$ \\
dx\_bin{:}t     & 0.01304  & 0.00329 & 15.68  & $7.5 \times 10^{-5}$ \\
\hline
\end{tabular}
\end{table}

\begin{table}[hbtp]
\caption{GEE Model Results (Simulated Data, Age Binned Without Practice Effect)}%
\centering
\begin{tabular}{lrrrr}
\hline
\textbf{Term} & \textbf{Estimate} & \textbf{Std. Error} & \textbf{Wald} & \textbf{p-value} \\
\hline
(Intercept)         & 0.09193  & 0.07428 & 1.53   & 0.2159 \\
age\_band\_kband2   & -0.00241 & 0.01746 & 0.02   & 0.8903 \\
age\_band\_kband3   & -0.04653 & 0.02128 & 4.78   & 0.0288 \\
age\_band\_kband4   & -0.09214 & 0.02349 & 15.39  & $8.8 \times 10^{-5}$ \\
age\_band\_kband5   & -0.11568 & 0.02780 & 17.31  & $3.2 \times 10^{-5}$ \\
dx\_bin             & -0.77973 & 0.02527 & 952.40 & $< 2 \times 10^{-16}$ \\
educ               & 0.09885  & 0.00503 & 385.76 & $< 2 \times 10^{-16}$ \\
gen                & 0.03588  & 0.02450 & 2.14   & 0.1431 \\
race\_lat          & -0.07673 & 0.00788 & 94.88  & $< 2 \times 10^{-16}$ \\
dx\_bin{:}t        & 0.00973  & 0.00300 & 10.29  & 0.0013 \\
\hline
\end{tabular}
\end{table}

\begin{table}[ptbh]
\caption{GEE Model Results (Simulated Data, With Practice Effect)}%
\centering
\begin{tabular}{lrrrr}
\hline
\textbf{Term} & \textbf{Estimate} & \textbf{Std. Error} & \textbf{Wald} & \textbf{p-value} \\
\hline
(Intercept)        & 0.28374 & 0.09455 & 9.01   & 0.00269 \\
age\_visit         & -0.00622 & 0.00182 & 11.65  & 0.00064 \\
dx\_bin            & -0.79729 & 0.02704 & 869.34 & $< 2 \times 10^{-16}$ \\
practiceprac1      & 0.20589  & 0.01082 & 362.13 & $< 2 \times 10^{-16}$ \\
practiceprac2      & 0.30321  & 0.01220 & 617.46 & $< 2 \times 10^{-16}$ \\
practiceprac3      & 0.38135  & 0.01414 & 727.41 & $< 2 \times 10^{-16}$ \\
practiceprac4      & 0.51345  & 0.01584 & 1050.84 & $< 2 \times 10^{-16}$ \\
practiceprac5      & 0.50425  & 0.01892 & 710.67 & $< 2 \times 10^{-16}$ \\
educ               & 0.09547  & 0.00535 & 318.25 & $< 2 \times 10^{-16}$ \\
gen                & 0.13455  & 0.02605 & 26.68  & $2.4 \times 10^{-7}$ \\
race\_lat          & -0.07779 & 0.00861 & 81.54  & $< 2 \times 10^{-16}$ \\
dx\_bin{:}t        & 0.01414  & 0.00450 & 9.86   & 0.00169 \\
\hline
\end{tabular}
\end{table}

\begin{figure}[htbp]
    \centering
    \begin{subfigure}[htbp]{0.49\linewidth}
        \centering
            \includegraphics[width=\linewidth]{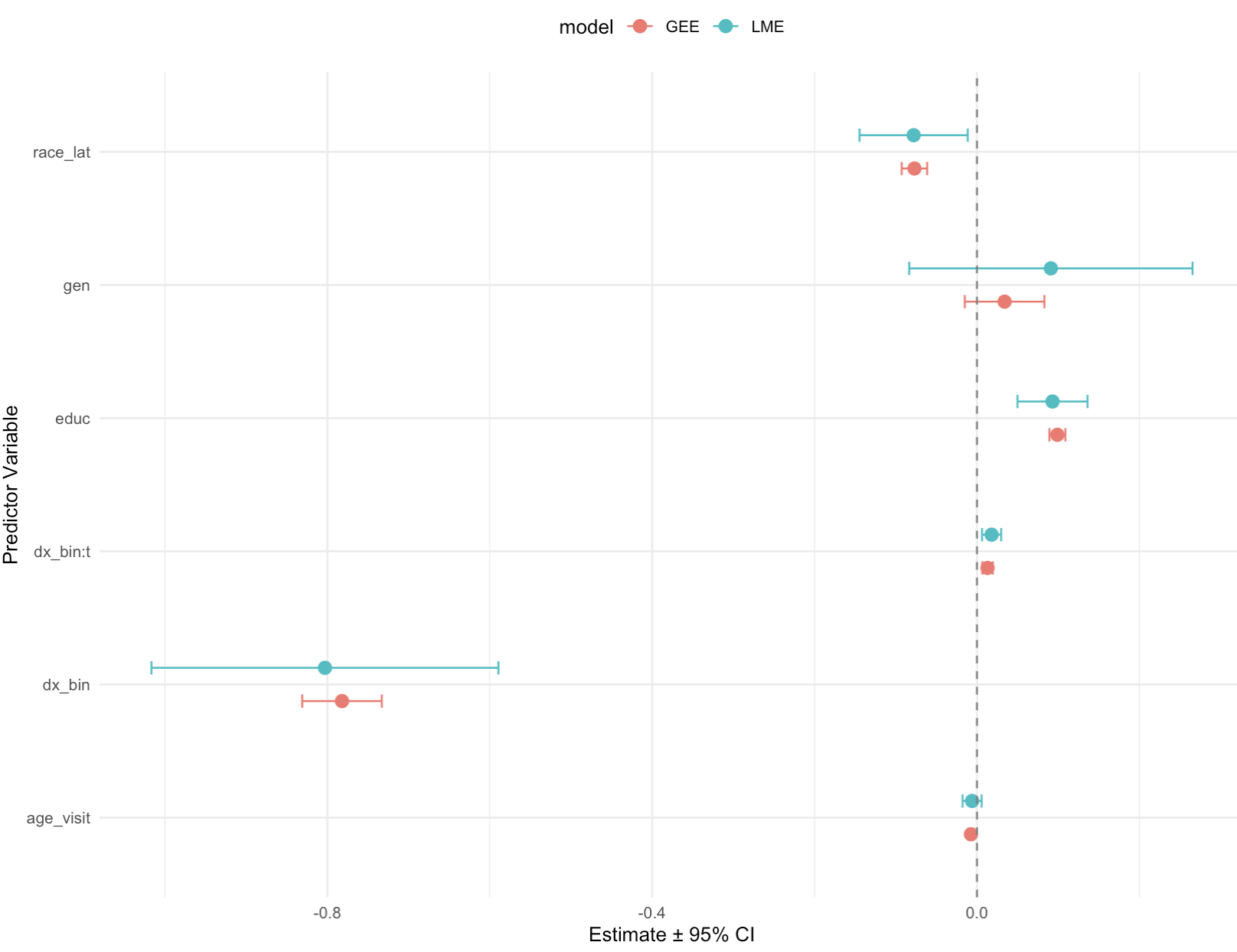}
        \caption{LME vs GEE Estimates (Without PE)}
    \end{subfigure}
    \hfill
    \begin{subfigure}[htbp]{0.49\linewidth}
        \centering
        \includegraphics[width=\linewidth]{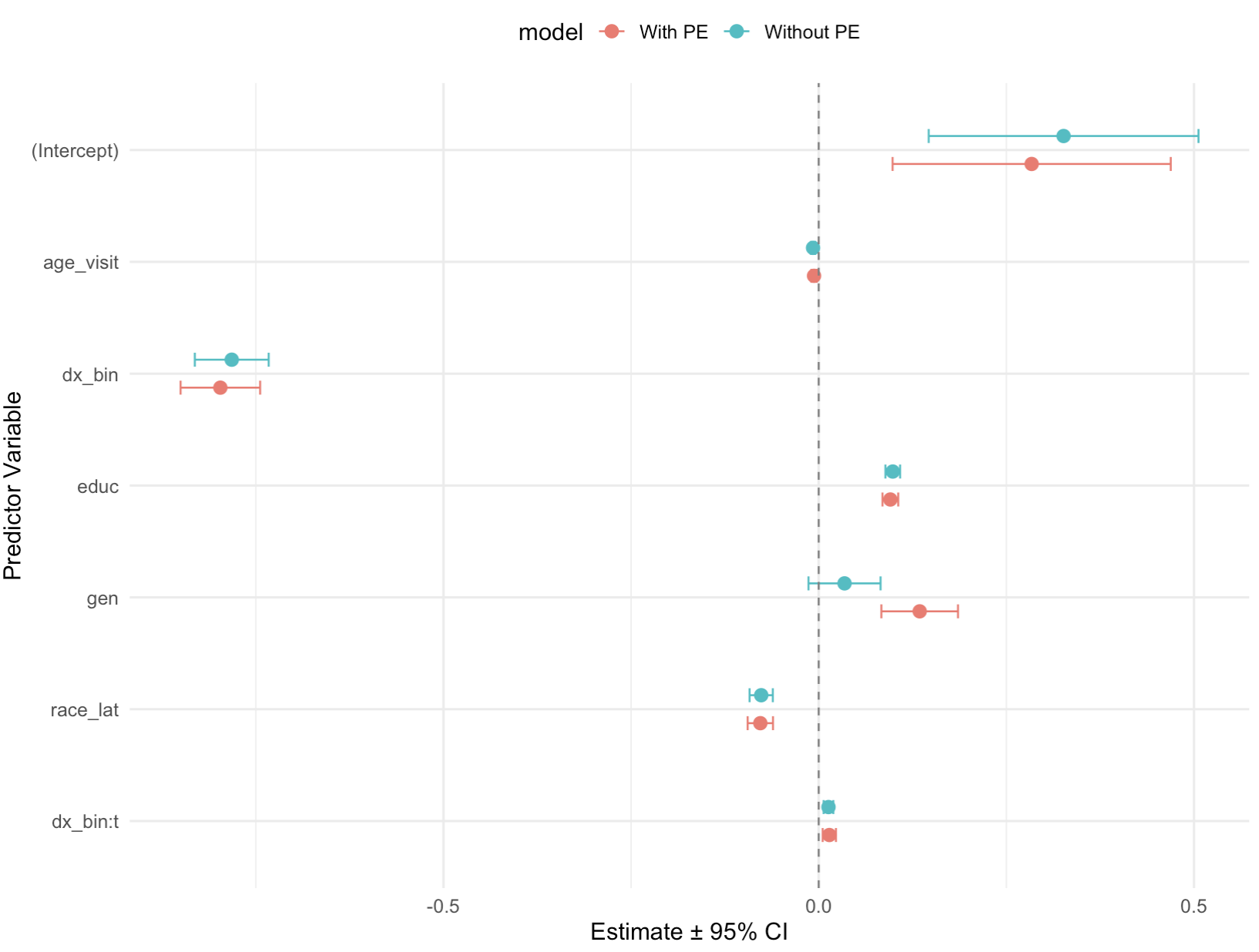}
        \caption{With vs Without PE Comparison}    \end{subfigure}
    \caption{Model comparison plots. Left: LME vs GEE. Right: With vs Without Practice Effects.}
    \label{fig:models_comparison}
\end{figure}
\begin{figure}
    \centering
    \includegraphics[width=0.49\linewidth]{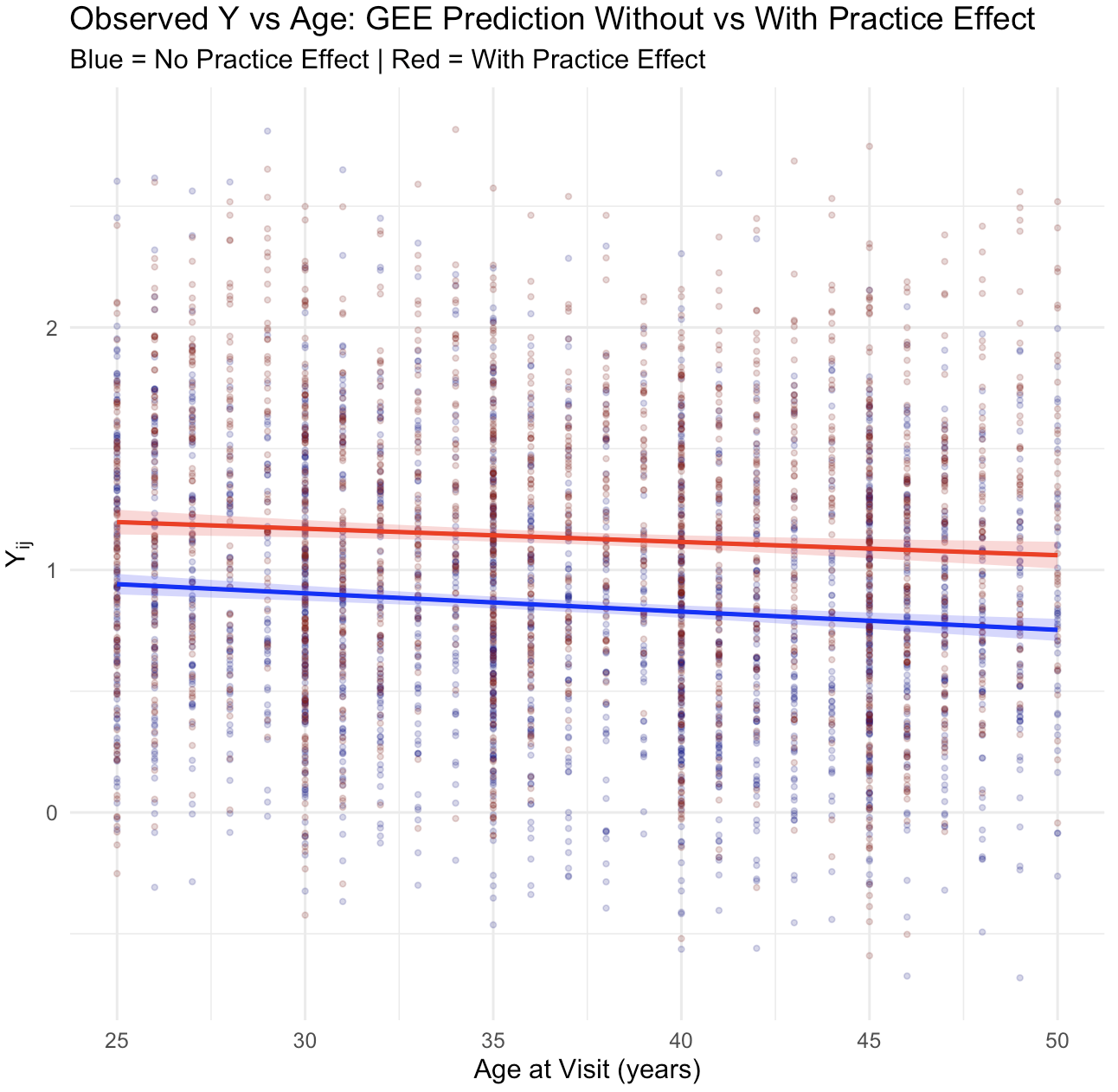}
    \caption{Plot of Outcome Y over Age at Visit}
\end{figure}

\newpage
\section{Discussion}
Longitudinal cognitive data reflect the combined influence of true temporal change and improvements driven by repeated test exposure. Extensive work has shown that practice effects (PE) can inflate performance, particularly early in follow-up, and may obscure or even reverse evidence of cognitive decline \cite{Salthouse2010a,Hausknecht2007,Goldberg2015}. In aging and psychiatric research, failure to account for PE risks mischaracterizing disease progression, treatment response, and group differences \cite{Rabbitt1993,Calamia2012}. The present study directly addressed this challenge by developing and evaluating a modeling framework that separates practice-related gains from aging-related change.

A central contribution of this work is the explicit specification of visit-level PE indicators, allowing practice gains to vary across assessments rather than assuming a constant or linear effect. This structure aligns with empirical patterns reported in neuropsychological cohorts, where the largest gains typically occur between the first and second assessments before plateauing \cite{Duff2012b,Bartels2010}. By estimating PE alongside demographic and diagnostic predictors, the model improves interpretability and reduces the risk that attenuated aging slopes are mistakenly attributed to preserved cognition.

The simulation findings further reinforce the importance of explicitly modeling PE. When practice effects were embedded in the data-generating process but omitted from analysis, aging-related decline was underestimated and diagnostic differences were muted—consistent with concerns raised in prior methodological evaluations \cite{McArdle2001,Wilson2002}. Incorporating PE restored the expected negative association between age and cognition and recovered group contrasts, demonstrating that conventional longitudinal models may yield misleading inferences when practice effects are ignored.

Another strength of the approach is the use of a flexible copula-based dependence structure to generate simulated data. This framework allows realistic within-subject correlation without imposing restrictive assumptions about temporal autocorrelation \cite{Fieberg2020,Nikoloulopoulos2018}. The use of both linear mixed-effects models and generalized estimating equations ensured that conclusions were not tied to a single analytic framework, consistent with recommendations for longitudinal cognitive research \cite{Fitzmaurice2012}.

Several limitations warrant consideration. Practice effects were modeled as monotonic and positive, yet PEs may diminish, reverse, or vary across cognitive domains, diagnostic groups, or testing intervals \cite{Salthouse2016}. Future work may evaluate spline-based, nonlinear, or domain-specific PE structures. The simulations did not incorporate additional real-world complexities such as alternate test forms, contextual influences, or informative dropout. Finally, the empirical dataset consisted of midlife adults from a single research program, potentially limiting generalizability to older populations or community-based samples.

Overall, these findings underscore the necessity of accounting for practice effects when interpreting longitudinal cognitive trajectories. The modeling strategy developed here offers a scalable, transparent, and empirically grounded approach for separating practice-related improvements from aging-related change. As repeated cognitive testing becomes increasingly common in clinical trials, epidemiologic studies, and preventative interventions, routine consideration of practice effects will be essential for accurate characterization of cognitive health across adulthood.

\end{document}